\documentclass[twocolumn,showpacs,preprintnumbers,amsmath,amssymb]{revtex4}
\usepackage{amsmath,amsfonts,diagbox,latexsym,amssymb,graphics,epsfig,subfigure,color,makeidx}
\usepackage{xcolor}
\usepackage[utf8]{inputenc}
\usepackage{graphicx,hyperref}
\usepackage{bm}
\usepackage{color}
\usepackage{multirow}

\begin{document}
	
	\title{Effects of magnetic fields on spinning test particles orbiting Kerr-Bertotti-Robinson black holes}
	\author{Yu-Kun Zhang$^{1,2,3}$,
		Shao-Wen Wei$^{1,2,3}$ \footnote{weishw@lzu.edu.cn,  corresponding author}
	}
	
	\affiliation{$^{1}$Key Laboratory of Quantum Theory and Applications of MoE, Gansu Provincial Research Center for Basic Disciplines of Quantum Physics, Lanzhou University, Lanzhou 730000, China\\
		$^{2}$Lanzhou Center for Theoretical Physics, Key Laboratory of Theoretical Physics of Gansu Province, School of Physical Science and Technology, Lanzhou University, Lanzhou 730000, People's Republic of China,\\
		$^{3}$Institute of Theoretical Physics $\&$ Research Center of Gravitation,
		Lanzhou University, Lanzhou 730000, People's Republic of China}
	
	\begin{abstract}
		In this paper, we study the kinematic effects of spinning test particles orbiting the Kerr-Bertotti-Robinson black hole. Employing with the Mathisson-Papapetrou-Dixon equations, we explore the dynamics of precessing orbits and distinct orbital types, including circular orbits and innermost stable circular orbits. Our results reveal the substantial impact of the magnetic field on the trajectories of spinning particles, particularly in regions characterized by significant radial distances. More importantly, our study shows that an augmented magnetic field necessitates an increased orbital angular momentum to uphold spinning particles within their characteristic orbits at equivalent radial distances. Our result contributes valuable insights to the understanding of the spinning celestial object motion around black holes endowed with magnetic fields.
	\end{abstract}
	
	\pacs{04.70.Bw, 04.25.-g, 04.50.-h}
	
	\maketitle
	
	\section{INTRODUCTION }
	
	In the realm of astrophysics, black holes, entities of profound observational and theoretical significance, typically exhibit distinctive rotational attributes \cite{Cui:2023uyb,Kerr:1963ud}. In particular, the vicinity of these cosmic giants often hosts magnetized plasma fields, detectable through modern astronomical instrumentation \cite{Blandford:1977ds,Bransgrove:2021heo}. Governed by the interplay of black hole gravity, rotation, and accretion dynamics, these magnetic fields serve as power source of fundamental physical phenomena such as the formation of black hole accretion disks and relativistic jets \cite{Narayan:2003by,Yang:2024kpz,Tchekhovskoy:2011zx,Hawley:2002aw,McKinney:2012vh,Blandford:1982xxl,Hawley:2005xs,Moscibrodzka:2015pda}.
	
	In the theoretical domain, the spacetime exterior to a rotating black hole is elegantly described by the Kerr metric---an exact solution to the Einstein field equations for an uncharged, rotating black hole, characterized by axial symmetry rather than spherical symmetry \cite{Kerr:1963ud}. However, the Kerr black hole model lacks provisions for external magnetic fields and thus falls short in the study of general magnetized astrophysical black holes. A recent breakthrough by Podolsky and Ovcharenko introduced the Kerr-Bertotti-Robinson (Kerr-BR) black hole \cite{Podolsky:2025tle}: a Kerr black hole enveloped in an asymptotically uniform external magnetic (or electric) field aligned with its rotational axis. This novel black hole is defined by three primary physical parameters: the black hole mass, its rotation, and the external magnetic field. Notably, when the magnetic field vanishes, the metric reduces to the familiar Kerr black hole. If the mass disappears, the metric corresponds to the conformally flat Bertotti-Robinson universe hosting a uniform Maxwell field. Moreover, when its rotation tends to zero, the spacetime transitions to a Schwarzschild black hole background immersed in a magnetic field.
	
	In contrast to Kerr-Melvin black holes that integrate a vertical magnetic field into the spacetime metric, categorized under Petrov type I, Kerr-BR black holes are classified under the Petrov type D class \cite{Podolsky:2025tle}. Distinctively, they exhibit bounded ergoregions and maintain electromagnetic fields that are both asymptotically finite and uniform. These characteristics makes Kerr-BR black holes a more realistic portrayal of a black hole immersed in an external electromagnetic field. Consequently, the Kerr-BR black hole solution finds extensive applications across diverse research domains, ranging from mathematical relativity to the field of relativistic astrophysics. The energy extraction via magnetic reconnection was examined in Ref. \cite{Zeng:2025olq}. The null geodesics is separable, while the time-like geodesics is not unless for the motion confined in the equatorial plane. Based on these results, the optical appearance and lensing were studied \cite{Wang:2025vsx,Zeng:2025tji,Wang:2025bjf,Astorino:2025lih,Ali:2025beh,Vachher:2025jsq}. All these results uncover that the magnetic field  has an important influence on these observable phenomena. More interestingly, the exact solution of the spinning charged black holes with nonaligned electromagnetic field are also obtained in Ref. \cite{Podolsky}.

	In our study, we deal with a system of profound theoretical interest and practical significance: the kinematic effects  of spinning test particles surrounding a Kerr-BR black hole. This investigation encapsulates the intricate motion of celestial entities in proximity to a Kerr-BR black hole. The fundamental framework for delineating the behavior of spinning test particles lies in the Mathisson-Papapetrou-Dixon (MPD) equations, originally documented in \cite{Mathisson:1937zz,Papapetrou:1951pa,Corinaldesi:1951pb,Dixon:1964cjb,Hojman:1976kn}. By making use of these equations, we ascertain the four-velocity and effective potential governing the trajectories of spinning particles around a Kerr-BR black hole. Our analysis emphasizes the impact of the magnetic field and the particle spin angular momentum on various orbital configurations, encompassing precessing orbits, circular orbits, and Innermost Stable Circular Orbits (ISCOs), which stand out for their paramount physical significance \cite{Suzuki:1997by,Shibata:1997mqo,Zdunik:2000qn,Baumgarte:2001ab,Marronetti:2003hx,Shahrear:2007zz,Hadar:2011vj,Akcay:2012ea,Hod:2013qaa,Chakraborty:2013kza,Hod:2014tpa,Zaslavskii:2014mqa,Isoyama:2014mja,Chartas:2016ckd,Harms:2016ctx,Lukes-Gerakopoulos:2017vkj,Chaverri-Miranda:2025fso,Du:2024ujg,Tan:2024hzw}.
	
	The manuscript is structured as follows: In Sect. \ref{Kbr}, we present an overview of the Kerr-BR black hole and MPD equations. Section \ref{effp} delves into the analysis of effective potentials and precessing orbits. The impact of the magnetic field on characteristic orbits, circular orbits, and ISCOs is investigated in Sect. \ref{cos}. Finally, in Sec. \ref{Conclusion}, we summarize and discuss our results. Throughout this work, we utilize $G=c=1$, and all physical quantities are nondimensionalized by setting the black hole mass and spinning particle mass to $1$.

	\section{Kerr-BR black hole and MPD equations}\label{Kbr}
	
	In this section, we shall briefly review the Kerr-BR black hole and MPD equations.
	
	\subsection{Kerr-BR black hole}
	
	Considering an asymptotically uniform external magnetic (or electric) field aligned with its rotational axis, the Kerr-BR black hole is described by the following line element
	\begin{eqnarray}
		ds^{2} &=\frac{1}{\Omega^{2}}\bigg[
		-\frac{Q}{\rho^{2}}\left(dt - a\sin^{2}\theta \, d\varphi\right)^{2} + \frac{\rho^{2}}{Q}dr^{2} + \frac{\rho^{2}}{P}d\theta^{2}  \nonumber \\
		&+ \frac{P}{\rho^{2}}\sin^{2}\theta \left(adt - \left(r^{2}+a^{2}\right)d\varphi\right)^{2}\bigg], \label{eq1}
	\end{eqnarray}
	where the metric functions read
	\begin{eqnarray}
		\rho^{2} &=& r^{2} + a^{2}\cos^{2}\theta,\label{eq2}\\
		P &=& 1 + B^{2}\left(M^{2}\frac{I_{2}}{I_{1}^{2}} - a^{2}\right)\cos^{2}\theta, \label{eq3}\\
		Q &=& \left(1 + B^{2}r^{2}\right)\Delta, \label{eq4}\\
		\Omega^{2} &=& \left(1 + B^{2}r^{2}\right) - B^{2}\Delta\cos^{2}\theta, \label{eq5}\\
		\Delta &=& \left(1 - B^{2}M^{2}\frac{I_{2}}{I_{1}^{2}}\right)r^{2} - 2M\frac{I_{2}}{I_{1}}r + a^{2}, \label{eq6}
	\end{eqnarray}
	with $I_{1} = 1 - \frac{1}{2}B^{2}a^{2}$ and $I_{2} = 1 - B^{2}a^{2}$. Here $M$ and $a$ are the mass and spin parameters of the black hole, while $B$ denotes the asymptotic value of the magnetic (or electric) field. Solving $\Delta=0$, one can obtain the radius of the black hole horizons \cite{Podolsky:2025tle}
	\begin{equation}
		r_{\pm} = \frac{M I_{2} \pm \sqrt{M^{2} I_{2} - a^{2} I_{1}^{2}}}{I_{1}^{2} - B^{2} M^{2} I_{2}} I_{1}.\label{eq7}
	\end{equation}
	Such property is similar to  the Kerr black hole, where two horizons are presented. Note that $r_+$ and $r_-$ not always corresponds to the outer and inner horizons of the black hole. With the increase of $B$, $r_-$ will be larger than $r_+$, and thus it acts as the outer horizon.
	
	Since we only consider the motion of particles on the equatorial plane, we adopt the induced metric with $\theta=\pi/2$
	\begin{eqnarray}
		g_{tt} &= \frac{a^2 - \overline{\Omega}^2 \Delta}{\overline{\Omega}^2 r^2}, \quad g^{tt} = \frac{a^2 \overline{\Omega}^2 \Delta - (r^2 + a^2)^2}{r^2 \Delta},\label{eq8} \\
		g_{rr} &= \frac{r^2}{\overline{\Omega}^4 \Delta}, \quad g_{t\varphi} = g_{\varphi t} = \frac{a\left( \overline{\Omega}^2 \Delta - r^2 - a^2 \right)}{\overline{\Omega}^2 r^2}, \label{eq9}\\
		g^{rr} &= \frac{\overline{\Omega}^4 \Delta}{r^2}, \quad g^{t\varphi} = g^{\varphi t} = \frac{a\left( \overline{\Omega}^2 \Delta - r^2 - a^2 \right)}{r^2 \Delta}, \label{eq10}\\
		g_{\varphi\varphi} &= \frac{(r^2 + a^2)^2 - a^2 \overline{\Omega}^2 \Delta}{\overline{\Omega}^2 r^2}, \quad g^{\varphi\varphi} = \frac{\overline{\Omega}^2 \Delta - a^2}{r^2 \Delta},\label{eq11}
	\end{eqnarray}
	where $\overline{\Omega}^2=1+B^2r^2$.

	\subsection{MPD equations}
	
	Now, let us turn to the MPD equations, which describe the motion of spinning test particles. According to Ref. \cite{Dixon:1964cjb}, these equations are given by
	\begin{eqnarray}
		\frac{DP^{\mu}}{D\lambda}&=-\frac{1}{2}R_{\nu \alpha \beta}^{\mu}u^{\nu}S^{\alpha \beta},\label{eq12}
		\\
		\frac{DS^{\mu \nu}}{D\lambda}&=P^{\mu}u^{\nu}-P^{\nu}u^{\mu},\label{eq13}
	\end{eqnarray}
	where $\lambda$ represents an arbitrary affine parameter, $R^\mu_{\nu\alpha\beta}$ signifies the Riemann curvature tensor, and $S^{\alpha\beta}$, $P^\mu$, and $u^\nu = dx^\nu/d\lambda$ denote the spin angular momentum tensor, four-momentum, and four-velocity of the spinning particle, respectively. Eqs. (\ref{eq12}) and (\ref{eq13}) contains 13 dynamical variables ($S^{\alpha\beta}$, $P^\mu$, and $u^\nu$), with the remaining undetermined degrees of freedom linked to the center-of-mass definition of spinning particles. In general relativity, the center of mass is observer-dependent \cite{Costa:2014nta}, requiring the introduction of a spin supplementary condition \cite{Mathisson:1937zz,Papapetrou:1951pa,Corinaldesi:1951pb,Dixon:1964cjb,1959Motion,Frenkel:1926zz,Ohashi:2003we,Kyrian:2007zz} to make the system fully determinate. In this paper, we adopt the Tulczyjew spin supplementary condition \cite{1959Motion}
	\begin{equation}
		P_\mu S^{\mu\nu} = 0.\label{eq14}
	\end{equation}
	The particle's mass and spin satisfy the following conservation relations respectively
	\begin{eqnarray}
		m^2 &=& -P^\mu P_\mu , \label{eq15}\\
		s^2 &=& \frac{1}{2}S^{\mu\nu}S_{\mu\nu} .\label{eq16}
	\end{eqnarray}
	To simplify our analysis, we will henceforth restrict our consideration to spinning particles moving in the equatorial plane of the black hole, with their spin vectors orthogonal to this plane. This assumption implies the following constraints
	\begin{equation}
		P^\theta = 0, \quad
		S^{\mu\theta} = 0.\label{eq17}
	\end{equation}
	After considering the spin supplementary condition Eq. (\ref{eq14}) and the constraints Eqs. (\ref{eq15}) and (\ref{eq16}), one has the non-vanishing components of the spin tensor
	\begin{eqnarray}
		S^{tr} &=& -\frac{s P_\phi}{m}\gamma^{-1} = -S^{rt}, \label{eq18}\\
		S^{\phi t} &=& -\frac{s P_r}{m}\gamma^{-1} = -S^{t\phi}, \label{eq19}\\
		S^{r\phi} &=& -\frac{s P_t}{m}\gamma^{-1} = -S^{\phi r},\label{eq20}
	\end{eqnarray}
	where $m$ represents the mass of the spinning particle, and $\gamma=\sqrt{-g_{tt}g_{rr}g_{\phi\phi}+g_{\phi t}g_{\phi t}g_{rr}}$.
	
	Furthermore, the existence of two Killing vectors $\xi^\mu = (\partial_t)^\mu$ and $\eta^\mu = (\partial_\phi)^\mu$ in the black hole spacetime enables the construction of two conserved quantities for spinning particles, namely the energy $e$ and the total angular momentum $j$
	\begin{eqnarray}
		e &=& -P_t - \frac{1}{2} \frac{sP_{\phi}}{m} \gamma^{-1} \partial_r g_{tt} +\frac{1}{2}\frac{sP_t}{m}\gamma^{-1} \partial_r g_{t\phi}, \label{eq21} \\
		j &=& P_{\phi}+\frac{1}{2}\frac{sP_{\phi}}{m} \gamma^{-1} \partial_r g_{t\phi}-\frac{1}{2}\frac{sP_t}{m}\gamma^{-1}\partial_r g_{\phi\phi}. \label{eq22}
	\end{eqnarray}
	The second steps in Eqs. (\ref{eq21}) and  (\ref{eq22}) are derived using the relations established in Eqs. (\ref{eq18}), (\ref{eq19}), and (\ref{eq20}). We now proceed to calculate the non-vanishing components of the momentum
	\begin{eqnarray}
		P^t &=& g^{tt}\frac{De-Bj}{DA-BC}+g^{t\phi}\frac{Ce-Aj}{CB-AD},\label{eq23}\\
		P^{\phi}&=& g^{\phi\phi}\frac{Ce-Aj}{CB-AD}+g^{\phi t}\frac{De-Bj}{DA-BC},\label{eq24}\\
		P^r &=&\sqrt{\frac{m^2+g_{tt}(P^t)^2+g_{\phi\phi}(p^{\phi})^2+2g_{\phi t}P^{\phi}P^t}{-g_{rr}}},\label{eq25}
	\end{eqnarray}
	where
	\begin{eqnarray}
		A &=& -1+\frac{1}{2}\frac{s \gamma^{-1}}{m}\partial_r g_{t\phi},\label{eq26}\\
		B &=& \frac{-1}{2}\frac{s \gamma^{-1}}{m}\partial_r g_{tt},\label{eq27}\\
		C &=& \frac{-1}{2}\frac{s \gamma^{-1}}{m}\partial_r g_{\phi\phi},\label{eq28}\\
		D &=& 1+\frac{1}{2}\frac{s \gamma^{-1}}{m}\partial_r g_{t\phi}\label{eq29}.
	\end{eqnarray}
	Combining with above result, the radial and angular velocities of the spinning particle can be obtain through (\ref{eq13})
	\begin{eqnarray}
		\frac{DS^{tr}}{D\lambda} &=& P^t u^r - P^r u^t, \label{eq30}\\
		\frac{DS^{t\phi}}{D\lambda} &=& P^t u^\phi - P^\phi u^t.\label{eq31}
	\end{eqnarray}
	Since the choice of affine parameter doesn't affect the particle's motion, for simplicity we choose $\lambda = t$, which leads to
	\begin{equation}
		u^t = \frac{dt}{dt} = 1, \quad
		u^r = \frac{dr}{dt} = \dot{r}, \quad
		u^\phi = \frac{d\phi}{dt} = \dot{\phi}.\label{eq32}
	\end{equation}
	Thus, from Eqs. (\ref{eq18}), (\ref{eq19}), (\ref{eq30}), and (\ref{eq31}), we have
	\begin{eqnarray}
		P^t u^r - P^r &=& \frac{s\gamma^{-1}}{2m} g_{\phi\phi} R_{\nu\alpha\beta}^{\phi} u^{\nu} S^{\alpha\beta},  \label{eq33}\\
		P^t u^\phi - P^\phi &=& -\frac{s\gamma^{-1}}{2m} g_{rr} R_{\nu\alpha\beta}^{r} u^\nu S^{\alpha\beta}.\label{eq34}
	\end{eqnarray}
	Consequently, the radial and angular velocities of the spinning particle are given by
	\begin{eqnarray}
		u^r = \dot{r} &=& \frac{P^r + \dfrac{s\gamma^{-1}}{2m} g_{\phi\phi} R_{t\alpha\beta}^{\phi} S^{\alpha\beta}}{P^t - \dfrac{s\gamma^{-1}}{2m} g_{\phi\phi} R_{r\alpha\beta}^{\phi} S^{\alpha\beta}},  \label{eq35}\\
		u^\phi =& \dot{\phi} &= \frac{P^\phi - \dfrac{s\gamma^{-1}}{2m} g_{rr} R_{t\alpha\beta}^{r} S^{\alpha\beta}}{P^t + \dfrac{s\gamma^{-1}}{2m} g_{rr} R_{\phi\alpha\beta}^{r} S^{\alpha\beta}}.\label{eq36}
	\end{eqnarray}
	On the other hand, it is known from previous studies that when MPD equations are used to describe particle orbits, the phenomenon of spacelike orbits may occur \cite{Tod:1976ud,Hojman:2013zz,Zalaquett:2013ifd,Zhang:2017nhl,Conde:2019juj,Toshmatov:2019bda,Nucamendi:2019qsn,Toshmatov:2020wky,Zhang:2020qew,Chen:2024sbc}. To ensure that the particle's trajectory is always timelike, we introduce a null constraint condition, namely the contraction of the four-velocity should be less than zero:
	\begin{eqnarray}
		u^\mu u_\mu &=& g_{tt} (u^t)^2 + g_{rr} (u^r)^2 + g_{\phi\phi} (u^\phi)^2 +2g_{t\phi}u^t u^{\phi} \nonumber \\
		&=& g_{tt} + g_{rr} (u^r)^2 + g_{\phi\phi} (u^\phi)^2 +2g_{t\phi}u^{\phi} < 0.\label{eq37}
	\end{eqnarray}
	Considering the motion confined to the equatorial plane of the black hole ($u^\theta = 0$), the emergence of spacelike orbits stems from MPD equations accounting only for the zeroth- and first-order terms in the multipole expansion. Including non-minimal spin-gravity coupling effects can eliminate such unphysical trajectories \cite{Deriglazov:2015zta,Deriglazov:2015wde,Ramirez:2017pmp,Deriglazov:2017jub}.
	
	Up to this point, we have derived physical quantities such as the radial and angular velocities of the spinning particle from MPD equations. To perform a detailed study on orbits of spinning particles, it is necessary for us to construct their effective potential. Noting that Eq. (\ref{eq35}) indicates that $u^r$ and $P^r$ are parallel, we can thus construct the effective potential using $P^r$ in place of $u^r$
	\begin{eqnarray}
		(P^r)^2 &=& \mathbb{A} e^2 +\mathbb{B} e + \mathbb{C} \nonumber \\
		&=&(e-\frac{-\mathbb{B}-\sqrt{\mathbb{B}^2-4\mathbb{A}\mathbb{C}}}{2\mathbb{A}}) \nonumber \\
		&\times&(e-\frac{-\mathbb{B}+\sqrt{\mathbb{B}^2-4\mathbb{A}\mathbb{C}}}{2\mathbb{A}}).\label{eq38}
	\end{eqnarray}
	The first step mentioned above can be easily accomplished. For the second step, the positive square root represents the four-momentum pointing to the future, while the negative square root represents it pointing to the past. Thus, one can chose the positive square root as the effective potential
	\begin{equation}
		V_{eff}=\frac{-\mathbb{B}+\sqrt{\mathbb{B}^2-4\mathbb{A}\mathbb{C}}}{2\mathbb{A}}.\label{eq39}
	\end{equation}
	It is worth noting that when the particle is spinless, i.e., when the spin parameter $s=0$ and the spin tensor $S^{\mu\nu}=0$, MPD equations reduce to the geodesic equations.

	\section{Effective potentials and precessing orbits}\label{effp}
	
	In this section, we attempt to examine the effective potential governing the motion of the spinning particles. Then the precessing orbits will be explored.
	
	\subsection{Effective potentials}
	
	We now proceed to discuss the effective potentials of spinning particles around Kerr-BR black hole, and bashed on which the subsequent study of precessing orbits of spinning particles will be given. Hereafter, the symbol $l$ denotes the orbital angular momentum of the spinning particle, and its relationship with $s$ and $j$ is given by $j=l+s$.
	
	\begin{figure*}[htbp]
		\center{\subfigure[$a=0,l=4,s=0$]{\label{fig1a}
				\includegraphics[width=6cm]{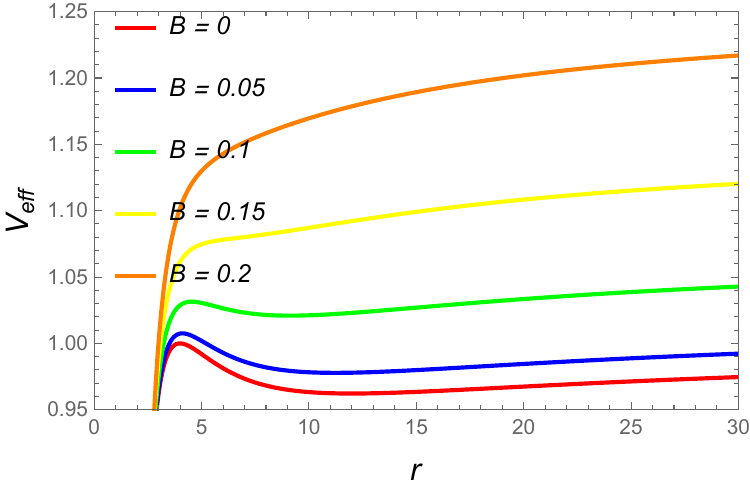}}
			\subfigure[$a=1,l=2,B=0$]{\label{fig1b}
				\includegraphics[width=6cm]{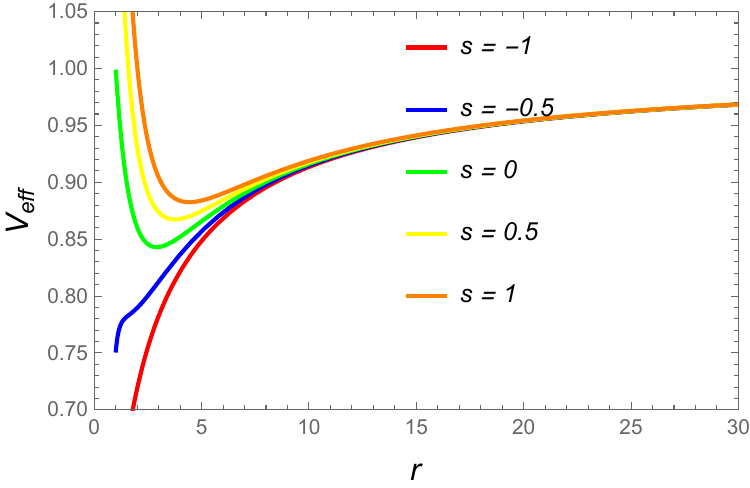}}
			\subfigure[$a=0.1,l=4,s=0.5$]{\label{fig1c}
				\includegraphics[width=6cm]{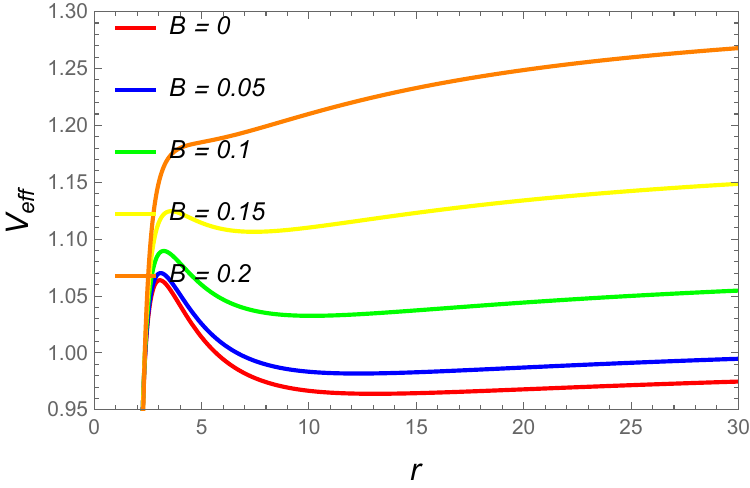}}
			\subfigure[$a=0.1,l=4,B=0.05$]{\label{fig1d}
				\includegraphics[width=6cm]{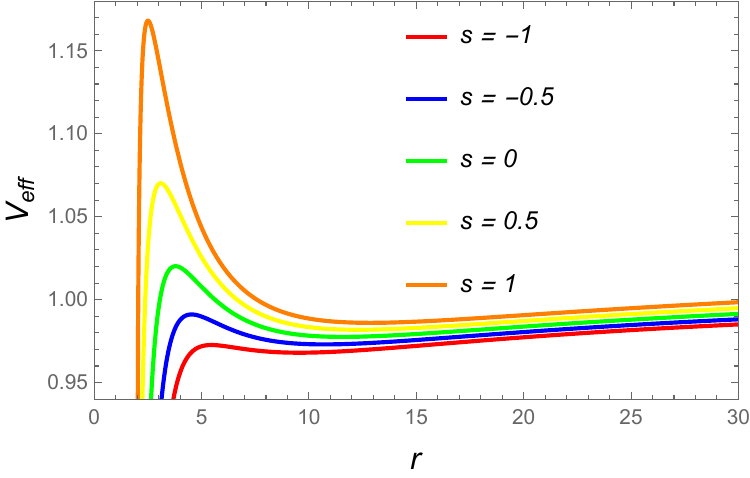}}}
		\caption{Effective potentials of spinning particles around Kerr-BR black hole. (a) $a=0$, $l=4$, $s=0$. (b) $a=1$, $l=2$, $B=0$. (c) $a=0.1$, $l=4$, $s=0.5$. (d) $a=0.1$, $l=4$, $B=0.05$.}
		\label{fig1}
	\end{figure*}
	
	In order to clearly show the effects of black hole parameters on the effective potential, we present its behaviors in Fig. \ref{fig1}. By varying the magnetic field $B$, the effective potential is described in Fig. \ref{fig1a}. In the vicinity of the black hole's event horizon, as the radial distance increases, the effective potential rises significantly, while exhibiting minimal sensitivity to variations in $B$. This reflects that gravity dominates the effective potential in regions with small radial distance. With increasing the radial distance, gravitational effects weaken, and effective potentials corresponding to different $B$ values rapidly disperse. Within a reasonable range of $B$, each effective potential curve features a maximum point, which corresponds to an unstable circular orbit.
	As radial distance further increases, a minimum point emerges in the effective potential which corresponds to a stable circular orbit. Obviously, for the sufficiently large radial distances, the effective potentials for different $B$ values do not converge to the same value. This indicates that in large radial distance regions, $B$ dominates the particle's trajectory. In Figs. \ref{fig1a} and \ref{fig1c}, one can find that, the effective potential increases with $B$. Moreover, the absence of a minimum point (i.e., the loss of stable orbits) becomes highly likely. This demonstrates that $B$ exerts a strong influence on the stability of particle orbits.
	
	When the black hole metric reduces to Kerr metric with $B=0$, our derived effective potential also accurately reproduces the Kerr metric-based effective potential (see Fig. \ref{fig1b}), which has been reported in Ref. \cite{Zhang:2017nhl}.
	
	In Fig. \ref{fig1c}, we show the behavior of the effect potential for both the nonvanishing black hole's rotation and the particle's spin, i.e., $a=0.1$ and $s=0.5$. Within a reasonable range of the magnetic field $B$ (from $0$ to $0.2$), the effective potential exhibits a variation similar to that given in Fig. \ref{fig1a}.
	
	As is presented in Fig. \ref{fig1d}, the spin angular momentum $s$ of the spinning particle also significantly modifies the profile of the effective potential. However, unlike in Fig. \ref{fig1b}, as radial distance increases, the effective potentials corresponding to different $s$ values fail to converge to the same magnitude once again. This further confirms that the magnetic field $B$ exerts a non-negligible influence on the motion of particles in large radial distance regions where gravitational effects are weak.
	
	\subsection{Precessing orbits}
	
	We now turn to explore the precessing orbits of the spinning particles around the Kerr-BR black hole.
	
\begin{figure*}[htbp]
		\center{\subfigure[$e=0.98,B=0,a=0.1,l=4,s=-0.5$]{\label{fig2a}
				\includegraphics[width=5cm]{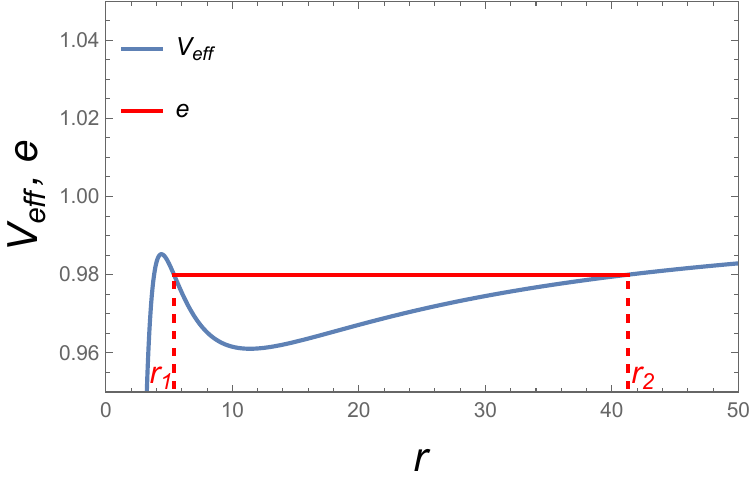}}
			\subfigure[$e=0.98,B=0.05,a=0.1,l=4,s=-0.5$]{\label{fig2b}
				\includegraphics[width=5cm]{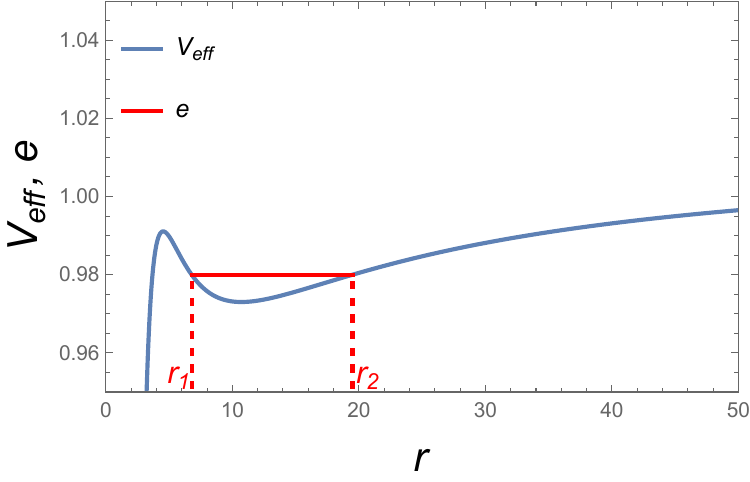}}
			\subfigure[$e=1.007,B=0.1,a=0.1,l=4,s=-0.5$]{\label{fig2c}
				\includegraphics[width=5cm]{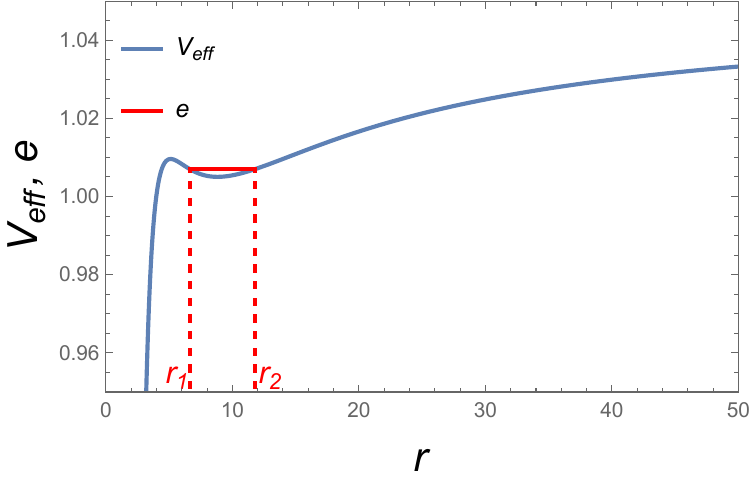}}
			\subfigure[$e=0.98,B=0,a=0.1,l=4,s=-0.5$]{\label{fig2d}
				\includegraphics[width=5cm]{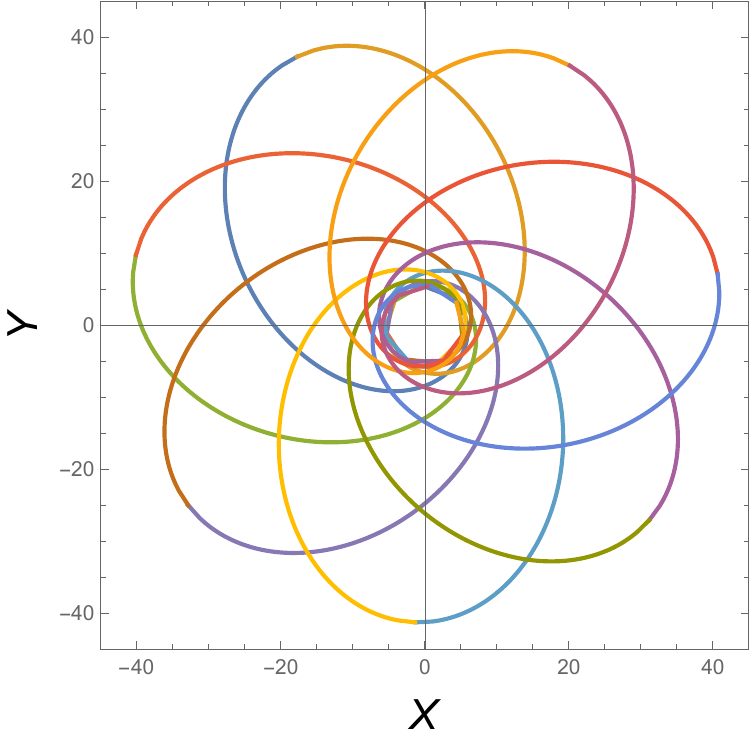}}
			\subfigure[$e=0.98,B=0.05,a=0.1,l=4,s=-0.5$]{\label{fig2e}
				\includegraphics[width=5cm]{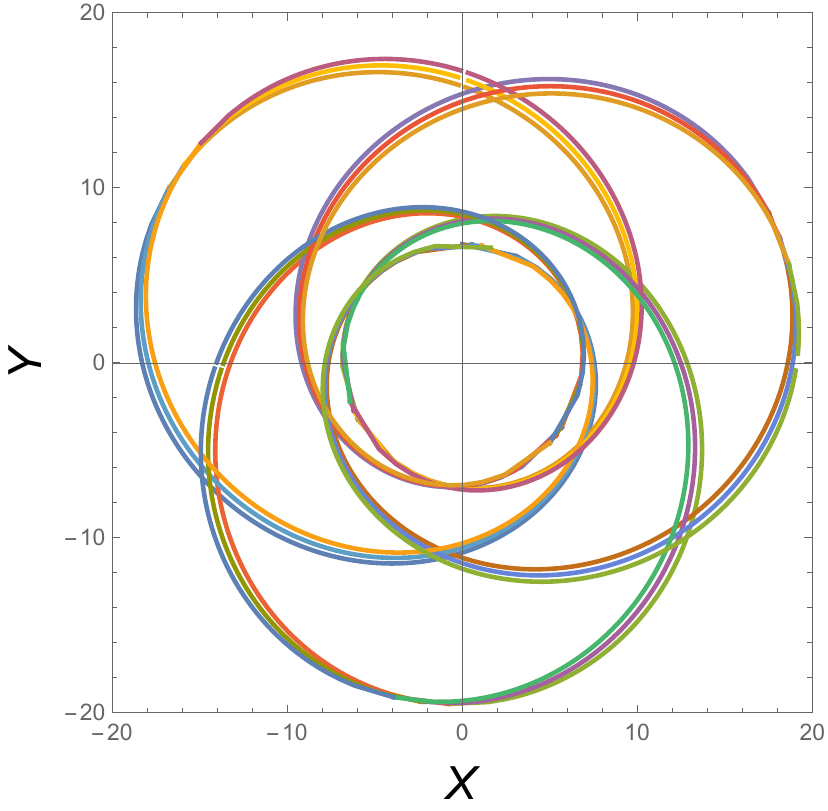}}
			\subfigure[$e=1.007,B=0.1,a=0.1,l=4,s=-0.5$]{\label{fig2f}
				\includegraphics[width=5cm]{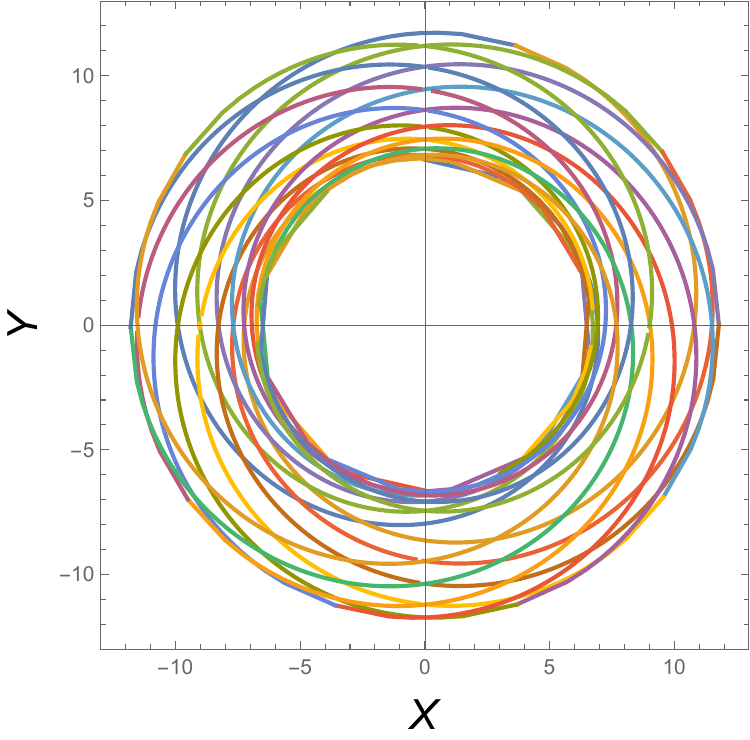}}
			\subfigure[$e=0.98,B=0,a=0.1,l=4,s=-0.5$]{\label{fig2g}
				\includegraphics[width=5cm]{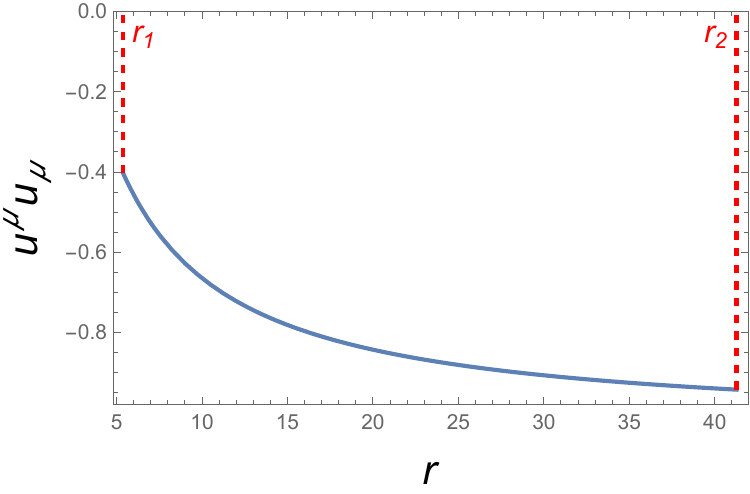}}
			\subfigure[$e=0.98,B=0.05,a=0.1,l=4,s=-0.5$]{\label{fig2h}
				\includegraphics[width=5cm]{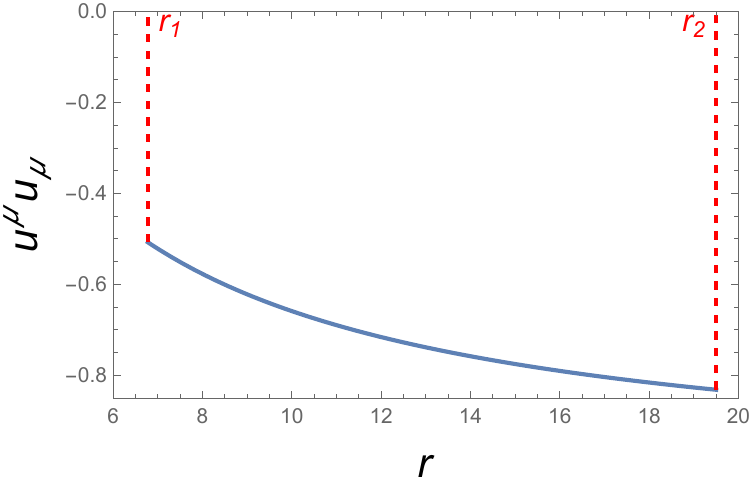}}
			\subfigure[$e=1.007,B=0.1,a=0.1,l=4,s=-0.5$]{\label{fig2i}
				\includegraphics[width=5cm]{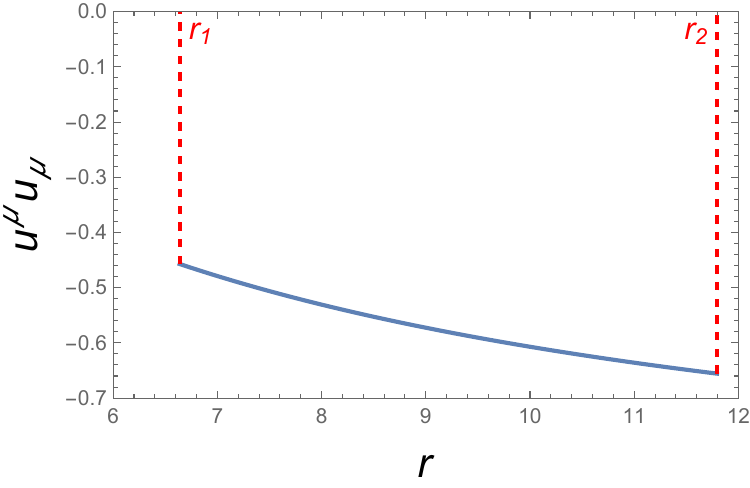}}}
		\caption{Precessing orbits of spinning particles around Kerr-BR black hole. The top row shows plots of the intersections between the effective potential and the energy $e$; the middle row presents plots of the precessing orbits of spinning particles; and the bottom row displays plots of the contraction of the particle's four-velocity within the radial range. Each column corresponds to the same parameter set.}
		\label{fig2}
	\end{figure*}

	According to Eqs. (\ref{eq35}) and (\ref{eq36}), the trajectory of spinning particles should be represented as a polar plot depicting the relationship between $r$ and $\phi$. We therefore eliminate $dt$ via dividing Eq. (\ref{eq36}) by Eq. (\ref{eq35}), obtaining the differential relation between $d\phi$ and $dr$
	\begin{equation}
		\frac{u^\phi}{u^r} = \frac{d\phi/dt}{dr/dt} \implies d\phi = \left(\frac{u^\phi}{u^r}\right) dr.\label{eq40}
	\end{equation}
	Having specified the particle's total energy $e$, which determines the radial range $r \in [r_1,r_2]$, we can now integrate the above expression:
	\begin{equation}
		\phi = \int_{r_1}^{r_2} \frac{u^\phi}{u^r}dr + \phi_0,\label{eq41}
	\end{equation}
	where $\phi_0$ is an arbitrarily specified initial angle, typically set to 0 for simplicity. From Figs. \ref{fig2a} and \ref{fig2b}, we can find that for fixed $e=0.98$, $a=0.1$, $l=4$, $s=-0.5$, $(r_1, r_2) \approx (5.4, 41.3)$ for $B=0$ and $(r_1, r_2) \approx (6.8, 19.5)$ for $B=0.05$. This uncovers the magnetic field significantly influence the particle motion in large distance. With the increase $B$, such range of $r_1$ and $r_2$ is further narrowed, see Fig. \ref{fig2c}.
	
	The above formula yields the particle's trajectory from $r_1$ to $r_2$, and using the same method, continuing to plot the trajectory from $r_2$ to $r_1$ gives the particle's motion over one complete cycle. Specifying multiple cycles produces trajectory plots, see Fig. \ref{fig2d}-\ref{fig2f}. Consistent with general relativity's prediction, the trajectories exhibit precession. Note that the radial distance first increases then decreases within each cycle, while the angle corresponding to each radial distance accumulates continuously from $r_1$ to $r_2$ and back to $r_1$; the same applies to subsequent cycles.
	
	The technical detail is that when $u^r$ approaches 0, the integral in Eq. (\ref{eq41}) seem to diverge. In order to avoid this computational difficulty, we introduce a smoothing factor $\epsilon \ll 1$ such that
	\begin{equation}
		\phi = \int_{r_1}^{r_2} \frac{u^\phi}{u^r + \epsilon}dr + \phi_0.\label{eq42}
	\end{equation}
	The specific order of the magnitude of $\epsilon$ depends on the divergence degree of $u^r$ in different cases. Theoretically speaking, smaller $\epsilon$ is better, but excessively small $\epsilon$ will cause it to fail to suppress divergence. On the other hand, when the particle's radial distance approaches $r_1$, even very small smoothing factors will cause observable mismatch between radial velocity and angular velocity, which will result in an angular difference between the plotted image and actual precession. Fortunately, if precise tracking of the relationship between particle's radial distance and angle is needed, we can always offset this difference through assignment of $\phi_0$.
	
	Next, we turn our attention back to precessing trajectories. Our objective is to investigate the precessing orbits of spinning particles under the same particle energy $e$  but different magnetic field $B$. However, as analyzed earlier, an increase in $B$ not only elevates the effective potential but also modifies its profile, even eliminating its minimum point entirely. Consequently, when $B=0.1$, there exists no energy $e$ of the same magnitude as those used for $B=0$ and $B=0.05$ that can define an effective potential interval containing a minimum point. It can be observed that even for the same energy $e$, different values of $B$ significantly affect the precessing angle. Although the precessing orbits are non-closed (rendering the concept of a ``period" inapplicable), we can intuitively estimate that the ``period" of the precessing orbit is approximately 7 and 3 for $B=0$ and 0.05, see Figs. \ref{fig2d} and \ref{fig2e}.
	
	On the other hand, based on prior study \cite{Tod:1976ud,Hojman:2013zz,Zalaquett:2013ifd,Zhang:2017nhl,Conde:2019juj,Toshmatov:2019bda,Nucamendi:2019qsn,Toshmatov:2020wky,Zhang:2020qew,Chen:2024sbc} , when describing spinning particle trajectories in certain black hole spacetimes, the MPD equations often leads to spacelike particle orbits with $u^{\mu}u_{\nu}>0$ in small radial distance regions with strong gravity. For Kerr-BR black hole, however, when $B$ and $s $ are within reasonable ranges, Figs. \ref{fig2g}, \ref{fig2h}, and \ref{fig2i} demonstrate that these whole orbits remain timelike with negative $u^{\mu}u_{\nu}$ throughout the radial region where particle precession occurs.

	\section{Characteristic orbits}\label{cos}
	
	Circular orbits of spinning particles, particularly the ISCOs, hold significant physical importance. In this section, we mainly focus on the effect of magnetic field on these two characteristic orbits for the spinning particles around a Kerr-BR black hole.
	
	\subsection{Circular Orbits}
	
	We first investigate the circular orbits of spinning particles, which is defined as
	\begin{equation}
		\frac{dV_{eff}}{dr} = 0.\label{eq43}
	\end{equation}
	If $\partial_r^2 V_{eff} \geq 0$, the circular orbit is stable; otherwise, the circular orbit is unstable.
	
	For fixed magnetic field $B$, the effective potential $V_{eff}$ is a function of parameters $(s, j, r)$. Therefore, to determine the point where the first derivative of the effective potential with respect to $r$ equals zero, we need to fix any two parameters and solve for the remaining one. The conventional approach is to fix $s$ and $j$ and solve for $r$ corresponding to $\partial_r V_{eff} = 0$. However, we found this would cause great inconvenience to the calculations. Therefore, we instead fix $s$ and $r$ and solve for $j$ corresponding to $\partial_r V_{eff} = 0$. To make the physical meaning of the plots clear, we use $l$ as the horizontal coordinate and $r$ as the vertical coordinate, with a specific $s$ value only serving as the parameter for its corresponding curve.
	
	\begin{figure*}[htbp]
		\center{\subfigure[$a=0.1,B=0.025$]{\label{fig3a}
				\includegraphics[width=6cm]{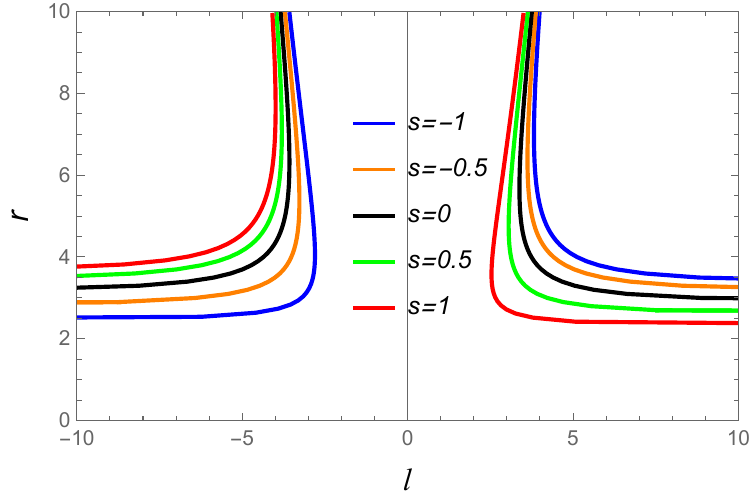}}
			\subfigure[$a=0.1,s=1$]{\label{fig3b}
				\includegraphics[width=6cm]{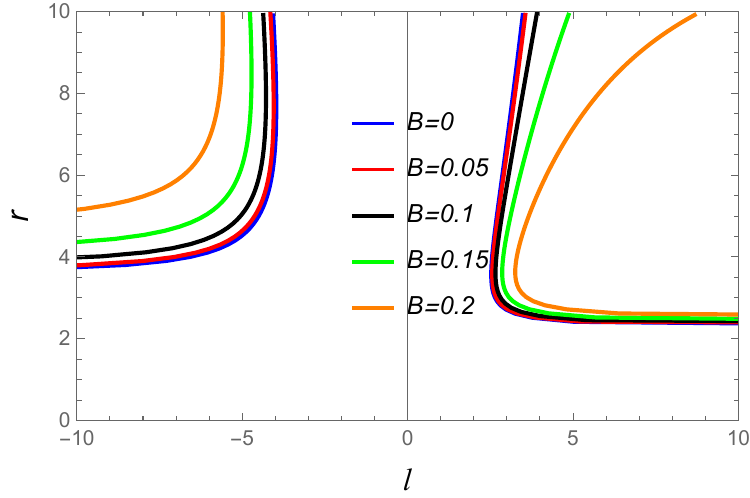}}}
		\caption{Circular orbits of the spinning particles around Kerr-BR black hole. (a) $a=0.1$, $B=0.025$. (b) $a=0.1$, $s=1$.}
		\label{fig3}
	\end{figure*}

	We plot the diagram of conventional circular orbits varying with the spin angular momentum $s$ of particles in Fig. \ref{fig3a}. As the radial distance approaches the event horizon where gravitational effects intensify, the angular velocity of the particle increases, leading to an increase in orbital angular momentum; as the radial distance moves away from the event horizon where gravitational effects weaken, the motion of the particle can always be approximately regarded as circular motion with an extremely large radius, and thus the particle has a large orbital angular momentum. The only point that needs to be emphasized is that, since Kerr-BR metric is not spherically symmetric, even when the spin angular momentum $s=0$, the two branches of circular orbits distributed on the left and right sides are not symmetric with respect to the vertical axis $r$.
	
	Similarly, when $s$ is fixed, we plot the diagram of circular orbits by varying with the magnetic field $B$ in Fig \ref{fig3b}. We only construct the circular orbit diagram for the case where $s$ takes positive values; it is expected that for the case where $s$ takes negative values, the diagram will exhibit similar behavior due to the symmetry. When the radial distance is close to the event horizon, Fig. \ref{fig3b} shows a trend similar to that of Fig. \ref{fig3a}, which once again indicates that the motion of particles in the small radial distance region is mainly affected by gravitational effects, while other parameters are relatively secondary. When the magnetic field $B$ is very small, its influence on the motion of particles is not obvious, which can also be seen from Figs. \ref{fig1a} and \ref{fig1c}; however, a slight increase in the magnetic field $B$ will exert a significant influence on the circular orbits of the particle. It can be observed that, especially for the right branch in Fig. \ref{fig3b}, an increase in the magnetic field $B$ obviously changes the circular orbits of particles in large radial distance regions, where particles need a larger orbital angular momentum to form circular orbits in the region with a larger magnetic field $B$. Further studies show that as the magnetic field $B$ continues to increase, the distribution of circular orbits will become disorganized, and even circular orbits cannot be solved directly at all. An increase in the spin angular momentum $s$ of the particle will also lead to similar results.

	\subsection{ISCO}
	
	The conditions for the stable circular orbits are
	\begin{equation}
		\frac{dV_{eff}}{dr}=0, \quad  \frac{d^2 V_{eff}}{dr^2}=0.\label{eq44}
	\end{equation}
	The so-called ``innermost" requires the radial distance of the circular orbit to be minimal. Integrating the two points of ``stability" and ``innermost", the ISCO should appear at the intersection of the curve where $\partial_r^2 V_{eff}$ first is less than 0 and then greater than 0 with the 0-axis, and for the physically meaningful ISCO, $\partial_r V_{eff}$ should be continuous.
	
	When one obtains an ISCO with fixed $B$ via (\ref{eq44}), a parameter set $(s_{ISCO},l_{ISCO},r_{ISCO})$ is determined. We can substitute this parameter set back into (\ref{eq39}) to obtain an effective potential with a definite numerical value
	\begin{equation}
		V_{eff}=e.\label{eq45}
	\end{equation}
	To sum up, the conditions for determining the ISCO are
	\begin{equation}
		V_{eff}=e, \quad \frac{dV_{eff}}{dr}=0, \quad \frac{d^2 V_{eff}}{dr^2}=0.\label{eq46}
	\end{equation}
	It is again emphasized that $V_{eff}=e$ originates from the parameter set $(s_{ISCO},l_{ISCO},r_{ISCO})$ obtained by solving $\partial_r V_{eff}=0$ and $\partial_r ^2 V_{eff}=0$, which is of great help for the subsequent discussion on the image of ISCOs.
	
	\begin{figure*}[htbp]
		\center{\subfigure[$a=0.1,B=0.05$]{\label{fig4a}
				\includegraphics[width=6cm]{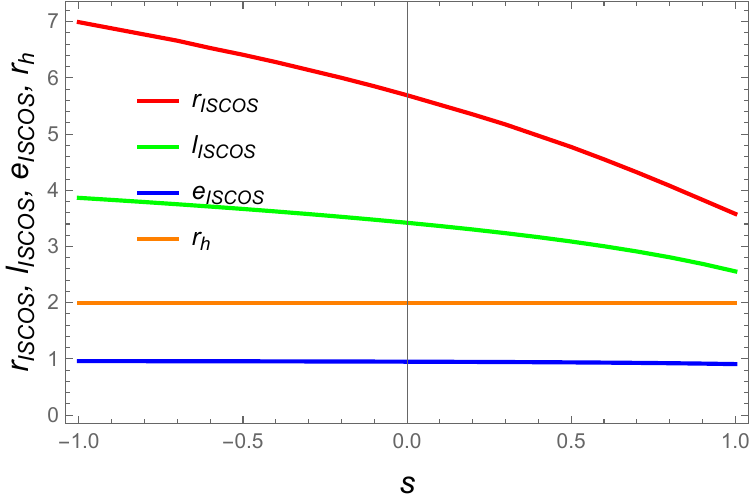}}
			\subfigure[$a=0.1,B=0.05$]{\label{fig4b}
				\includegraphics[width=6cm]{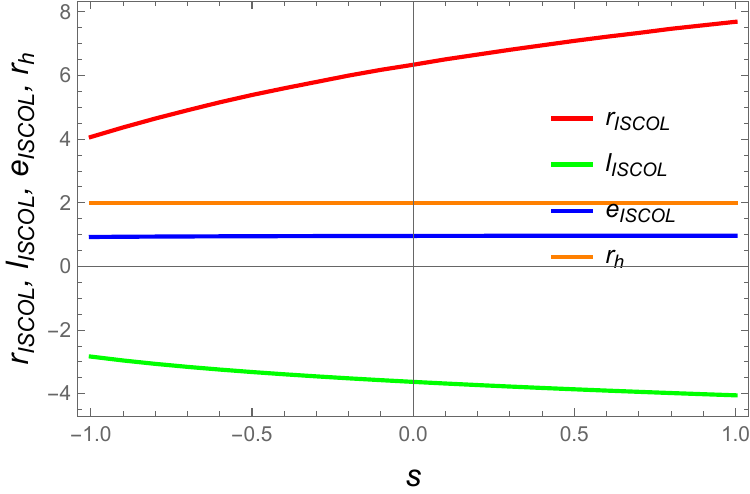}}
			\subfigure[$a=0.1,s=0.5$]{\label{fig4c}
				\includegraphics[width=6cm]{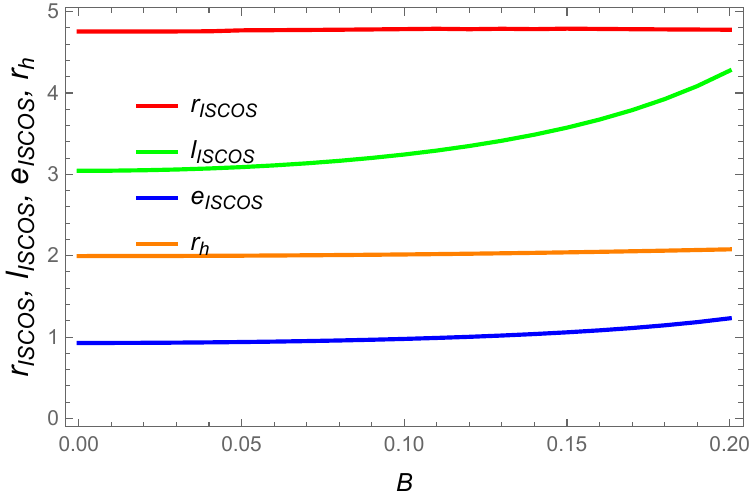}}
			\subfigure[$a=0.1,s=0.5$]{\label{fig4d}
				\includegraphics[width=6cm]{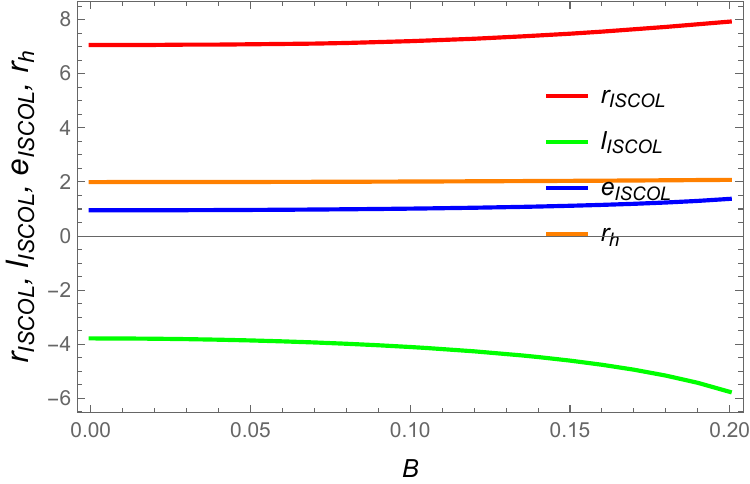}}}
		\caption{ISCOs (ISCOS/ISCOL) of the spinning particles around a Kerr-BR black hole. (a) ISCOS with $a=0.1$ and $B=0.05$. (b) ISCOL with $a=0.1$ and $B=0.05$. (c) ISCOS with $a=0.1$ and $s=0.5$. (d) ISCOL with $a=0.1$ and $s=0.5$.}
		\label{fig4}
	\end{figure*}
	
	For clarity, we will denote the ISCO on the left branch of circular orbit diagrams as ISCOL and the one on the right branch as ISCOS. Analysis of Figs. \ref{fig4a}-\ref{fig4d} reveals a similarity in the general trend of ISCO parameter sets with spin angular momentum $s$ to previous standard studies \cite{Zhang:2017nhl,Chen:2024sbc,Zhang:2020qew}. It is worth noting that by deriving Eq. (\ref{eq45}) for each specific set of $(s_{ISCO},l_{ISCO},r_{ISCO})$, an expanded data set $(e_{ISCO},s_{ISCO},l_{ISCO},r_{ISCO})$ was obtained. In ISCO plots, the $e$ values differ across various expanded data sets. Each expanded data set in the ISCO plots is represented by intersection points between a vertical line and the four curves, leading to non-horizontal $e$ lines. A key distinction arises due to the non-spherical metric symmetry, resulting in the loss of parameter space symmetry between ISCOS and ISCOL. Direct derivation of ISCOL from ISCOS is not feasible due to this metric asymmetry
	\begin{equation}
		\begin{aligned}
			&s_{ISCOL}=-s_{ISCOS}, \quad l_{ISCOL}=-l_{ISCOS}, \\ &r_{ISCOL}=r_{ISCOS}, \quad e_{ISCOL}=e_{ISCOS}.\label{eq47}
		\end{aligned}	
	\end{equation}
	
	Changes in the spin angular momentum have a significant impact on $r_{ISCO}$, whereas changes in the magnetic field have a relatively minor impact on $r_{ISCO}$. As observed in Figs. \ref{fig4a} and \ref{fig4b}, when $s$ and $l$ switch from being antiparallel to parallel, both $r_{ISCO}$ and $l_{ISCO}$ decrease noticeably, which is corresponding to an ISCO that is closer to the event horizon. For Figs. \ref{fig4c} and \ref{fig4d}, however, as $B$ increases, while $l_{ISCO}$ still increases noticeably, $r_{ISCO}$ changes very slightly. In other words, for spinning particles at the same $r_{ISCO}$, when they are in a magnetic field, they require a larger $l_{ISCO}$ to remain in that orbit compared to the case where no magnetic field exists. This is consistent with the conclusion we derived from our analysis of circular orbit diagrams.
	
	Given the meticulous selection of parameters $s$ and $B$ to ensure solvability for circular orbits (and subsequently, ISCOs), our analysis reveals that all scenarios depicted in the aforementioned four figures satisfactorily meet the timelike condition, with no observation of spacelike ISCOs. Nevertheless, broadening the range of $s$ or $B$ increases the likelihood of encountering situations where no circular orbits can be resolved at all.

	\section{ Summary and conclusions }\label{Conclusion}
	
	In this paper, we studied the motion trajectories of spinning particles around Kerr-BR black hole using the effective potential analysis method. We considered particles moving within the equatorial plane of the black hole, with their spin angular momentum aligned either parallel or antiparallel to their orbital angular momentum. Given the minimal spin-gravity coupling effect, the traditional description of particle motion using geodesic equations proves inadequate, prompting the adoption of MPD equations. Ensuring the theory's physical integrity, we meticulously uphold the null constraint on trajectories. Our results reveal that the external magnetic field $B$ surrounding the Kerr-BR black hole exerts a substantial influence on the motion of spinning particles, particularly in regions of large radial distances.
	
	We began by introducing the metric of Kerr-BR black hole and MPD equations, which are used to construct the basic physical quantities required for the study, primarily including the effective potential and four-velocity of the spinning particle. Our study reveals that in small radial distance regions, the influence of the magnetic field $B$ on the particle's effective potential is nearly negligible, as the dominant effect comes from the strong gravitational field. However, as the radial distance increases, $B$ rapidly modifies the profile of the effective potential of the spinning particle. Furthermore, with a further increase in radial distance, effective potentials corresponding to different values of $B$ do not converge, exhibiting significant differences. Notably, a sufficiently large $B$ can easily lead to the absence of a minimum point in the effective potential which implies the non-existence of stable orbits.
	
	Through the fixation of the total energy of the spinning particle, we explored its precessing orbits around the Kerr-BR black hole. Our observations unveil that even minor fluctuations in $B$ can yield notable disparities in the precessing angles, offering an indirect approach for determining the value of $B$. Different vales of $B$ lead to distinct profiles of the effective potential, such that the total energy of a single particle often fails to cover all scenarios, which is consistent with the aforementioned energy differences induced by $B$. Furthermore, within a suitable range of $B$, the timelike nature of the precessing orbits is consistently upheld.
	
	Then we turned to study the circular orbits of spinning particles around Kerr-BR black hole. We found that, due to the non-spherical symmetry of the metric, the two branches of the circular orbits are not symmetric with respect to the vertical axis $r$ even when the particle's spin angular momentum is zero. For different values of $B$, in small radial distance regions close to the black hole's event horizon, the asymptotic behavior of circular orbits is similar to that in conventional black holes, where gravity is the primary factor influencing the formation of circular orbits. As radial distance increases, gravitational effects weaken, and $B$ quickly becomes the dominant factor shaping circular orbits. This is manifested in the requirement for a larger orbital angular momentum to maintain a circular orbit at the same radial distance compared to the cases of $B=0$ or extremely small $B$. A sufficiently large $B$ can directly lead to the non-existence of circular orbits.
	
	At the end we examined the ISCOs of the spinning particles around a Kerr-BR black hole. The result shows that, due to the non-spherical symmetry of the metric, the profile of ISCOL cannot be derived by simply swapping the particle's spin angular momentum and orbital angular momentum from ISCOS. Importantly, as $B$ increases, to maintain the ISCO at the same radial distance, spinning particles require a larger $l_{ISCO}$ than that when $B$ is absent. Such a scenario does not occur for a varying $s$. Within an appropriate range of $s$ and $B$, the timelike nature of ISCOs is well guaranteed. Notably, a sufficiently large $s$ or $B$ more likely leads directly to the non-existence of circular orbits (and thus the non-existence of ISCOs).
	
	Our conclusions point to one key insight that the magnetic field $B$ of Kerr-BR black hole exerts a significant influence on particle motion in regions with large radial distance. Incidentally, in Kerr-BR black hole background, both $B$ and $s$ when within reasonable ranges effectively ensure the timelike nature of particle trajectories. In future investigations, we aim to analyze how $B$ impacts other dynamic aspects of particles, such as periodic orbits or spherical orbits, in regions characterized by significant radial distances.

	\acknowledgments
We would like to thank Dr. Yu-Peng Zhang for useful discussions. This work was supported by the National Natural Science Foundation of China (Grants No. 12475055, and No. 12247101), the Fundamental Research Funds for the Central Universities (Grant No. lzujbky-2025-jdzx07), and the Natural Science Foundation of Gansu Province (No. 22JR5RA389, No.25JRRA799).


\end{document}